\documentclass[usenatbib]{mn2e}
\usepackage{graphicx}
\def\sigmat{\sigma_\mathrm{T}}
\def\me{m_\mathrm{e}}
\def\mpr{m_\mathrm{p}}
\def\ms{M_\odot}
\def\ledd{L_\mathrm{Edd}}
\def\lcrit{L_\mathrm{crit}}
\def\tc{T_\mathrm{C}}
\def\teq{T_\mathrm{eq}}
\def\mbh{M_\mathrm{BH}}
\def\mcrit{M_\mathrm{\rm BH, crit}}
\def\tvir{T_\mathrm{vir}}
\def\rc{R_\mathrm{C}}
\def\mgas{M_\mathrm{gas}}

\def\mstar{M_*}
\def\rbh{R_{\rm BH}}
\def\rbon{R_{\rm B}}
\def\ncrit{n_{\rm crit}}
\def\tflow{t_{\rm dyn}}
\def\reff{R_{\rm e}}
\def\etaesc{\eta_{\rm esc}}

\def\lesssim{\hbox{\rlap{\hbox{\lower4pt\hbox{$\sim$}}}\hbox{$<$}}}
\def\gtrsim{\mathrel{\hbox{\rlap{\hbox{\lower4pt\hbox{$\sim$}}}\hbox{$>$}}}}
\def\beq{\begin{equation}}
\def\eeq{\end{equation}}
\def\beqa{\begin{eqnarray}}
\def\eeqa{\end{eqnarray}}
\def\Msun{M_{\odot}}
\def\yrs{{\rm yr}}
\def\mp{m_p}

\def\ergs{{\rm erg/s}}
\def\ergscm{{\rm erg\, cm^3/s}}
\def\Reff{R_{\rm e}}
\def\Tvir{T_{\rm vir}}
\def\vcirc{v_c}
\def\DEgas{\dot E}
\def\DEgasC{\dot E_C}

\def\DEgasSNw{\dot E_{H,SN}}
\def\DEgasII{\dot E_{II}}
\def\DEgasIa{\dot E_{Ia}}
\def\DEgasAGN{\dot E_{H,AGN}}
\def\DEgasAD{\dot E_{ad}}
\def\DEgasRW{\dot E_{H,w}}

\def\Mgal{M_{gal}}

\def\Minf{M_{inf}}
\def\DMinf{\dot\Minf}
\def\tinf{\tau_{inf}}

\def\Mgas{M_{gas}}
\def\DMgas{\dot\Mgas}

\def\LT{\Lambda (T)}
\def\rhom{\bar\rho_e}
\def\rhobon{\rho_{\rm B}}

\def\Mrec{M_{rec}}
\def\DMrec{\dot\Mrec}

\def\Mstar{M_*}
\def\DMstar{\dot\Mstar}
\def\Wstar{W_*}
\def\WIa{W_{Ia}}
\def\alstar{\alpha_*}
\def\tdyn{\tau_{dyn}}
\def\tcool{\tau_{cool}}
\def\theat{t_{\rm heat}}
\def\tflow{t_{\rm dyn}}
\def\theatSNw{\tau_{heat,SN,w}}

\def\tz{\tau_0}
\def\tIa{\tau_{Ia}}
\def\Rstar{R_*}
\def\destar{\delta_*}
\def\deIa{\delta_{Ia}}
\def\Mstari{M_{*,i}}

\def\MII{M_{II}}
\def\etasn{\eta_{SN}}

\def\Esn{E_{SN}}
\def\NII{N_{II}}
\def\NIa{N_{Ia}}
\def\thetaIa{\vartheta_{Ia}}

\def\Mesc{M_{esc}}
\def\DMesc{\dot\Mesc}
\def\tesc{\tau_{esc}}
\def\vesc{v_{esc}}
\def\etaesc{\eta_{\rm esc}}

\def\Mbh{M_{BH}}
\def\DMbh{\dot\Mbh}
\def\DMbhac{\dot M_{BH,acc}}

\def\bhstar{\beta_{BH,*}}
\def\DMedd{\dot M_{\rm Edd}}
\def\DMbon{\dot M_{\rm B}}
\def\csound{c_{\rm s}}
\def\rbon{R_{\rm B}}
\def\fed{f_{\rm Edd}}
\def\Ledd{L_{\rm Edd}}

\def\ncrit{n_{\rm crit}}
\def\Lcrit{L_{\rm crit}}
\def\rbh{R_{\rm BH}}

\title[Radiative feedback from quasars and growth of massive black holes 
in spheroids]{Radiative feedback from quasars and the growth
of massive black holes in stellar spheroids}

\author[S. Yu. Sazonov et al.]{S. Yu. Sazonov$^{1,2}$\thanks{E-mail:
sazonov@mpa-garching.mpg.de}, J. P. Ostriker$^{3}$, L. Ciotti$^4$, 
and R. A. Sunyaev$^{1,2}$\\
$^{1}$Max-Planck-Institut f\"ur Astrophysik,
Karl-Schwarzschild-Str. 1, 85740 Garching bei M\"unchen, Germany\\
$^{2}$Space Research Institute, Russian Academy of Sciences,
Profsoyuznaya 84/32, 117997 Moscow, Russia\\ 
$^3$Institute of Astronomy, Madingley Road, CB3 0HA Cambridge\\
$^4$Department of Astronomy, University of Bologna, via Ranzani 1,
I-40127 Bologna, Italy}

\begin{document}


\maketitle


\begin{abstract}
We discuss the importance of feedback via photoionization and Compton
heating on the co-evolution of massive black holes (MBHs) at the
center of spheroidal galaxies and their stellar and gaseous
components. We first assess the energetics of the radiative feedback
from a typical quasar on the ambient interstellar medium. We then
demonstrate that the observed $\mbh$--$\sigma$ relation could be
established following the conversion of most of the gas of an
elliptical progenitor into stars, specifically when the gas-to-stars
mass ratio in the central regions has dropped to a low level
$\sim$0.01 or less, so that gas cooling is no longer able to keep up
with the radiative heating by the growing central MBH. A considerable
amount of the remaining gas will be expelled and both MBH accretion and star
formation will proceed at significantly reduced rates thereafter, in
agreement with observations of present day ellipticals. We find
further support for this scenario by evolving over an equivalent
Hubble time a simple, physically based toy model that additionally
takes into account the mass and energy return for the spheroid
evolving stellar population, a physical ingredient often neglected in
similar approaches.
\end{abstract}

\section{Introduction}
\label{intro}
 
Elliptical galaxies are, with few exceptions, poor with respect to
interstellar gas \citep[e.g.][]{ofp01}. Also, elliptical galaxies
invariably contain central massive black holes (MBHs), and there
exists a tight relationship between the characteristic stellar
velocity dispersion $\sigma$ and the black hole mass $\mbh$
\citep{fm00,tgb+02}, and between $\mbh$ and the host spheroid mass in
stars, $\Mstar$ \citep{mtr+98}. Are these two facts related? Here we
focus on a scenario in which the mass of the central MBH grows within
gas rich elliptical progenitors until star formation has reduced the
gas fraction in the central regions to of order 1 per cent of the
stellar mass. Then radiative feedback, during episodes when the
luminosity from the central MBH approaches its Eddington limit, heats
and drives much of the remaining gas out of the galaxy, limiting both future
growth of the MBH and future star formation to low levels.

We show that for a typical quasar spectral energy distribution (SED)
the limit on the central MBH produced by this argument coincides
accurately with the observed $\mbh$-$\sigma$ relation and we predict
that the observed power law should break down below $10^4 \ms$ and
above a few $10^9\ms$. In our calculations, we adopt the average
quasar SED derived by \citet[][ hereafter SOS]{sos04} from cosmic
background radiation fields supplemented by information from
individual objects. Of key importance is that the UV and high energy
radiation from {\sl a typical quasar can photoionize and heat a low density
gas up to an equilibrium Compton temperature ($\tc\approx 2\times
10^7$\,K) that exceeds the virial temperatures of giant ellipticals};
the radiative effects on cluster gas are expected to be minimal
\citep[e.g.][]{cop04} but the effects on gas within
the host galaxy can be substantial.  We note that the present paper is
complementary to those by \citet[][ hereafter CO97,01]{co97,co01} in
that, while it does not attempt to model the complex flaring behavior
of an accreting MBH with an efficient hydrodynamical code, it does do
a far more precise job of specifying the input spectrum and the
detailed atomic physics required to understand the interaction between
that spectrum and the ambient interstellar gas in elliptical galaxies.

The evolutionary scenario considered could explain several key
observational facts. First, that giant ellipticals are old -- they end
their period of star formation early in cosmic time, since the
radiative output from the central MBHs limits (in cooperation with the
energetic input due to star formation) the gas content to be at levels
for which ongoing star formation is minimal. Secondly, gas rich,
actively star forming galaxies at redshift $z\sim 3$, including Lyman
Break Galaxies and bright submillimeter SCUBA galaxies, generally
exhibit only moderate active galactic nucleus (AGN) activity
\citep{s+02,a+03,l+04}, indicating that their central MBHs are still
growing. This suggests that the formation of a spheroid probably
closely preceeds a quasar shining phase, which finds further support
in spectroscopic observations showing that quasars live in a
metal-enriched environment (e.g. \citealt{hf99}). The redshift
evolution of the quasar emissivity and of the star formation history
in spheroids is thus expected to be roughly parallel since $z\sim 3$,
which also is consistent with observations
(e.g. \citealt{hco04,hkb+04}).

We note that many authors have already recognized the importance of
feedback as a key ingredient of the mutual MBH and galaxy evolution
(among them \citealt{bt95}; CO97; \citealt{sr98,f99}; CO01;
\citealt{bs01,cv01,gsm+01,cv02,k03,wl03,hco04,gds+04,mqt04,sdh04}). 
What is new about this work is the stress on one component of the
problem that has had relatively little attention -- the {\it radiative
output} of the central MBH. 

We emphasize that the radiative output is not the only, nor even
necessarily the dominant mechanism whereby feedback from accretion
onto central MBHs can heat gas in elliptical galaxies. \citet{bt95}
have stressed that the mechanical input from radio jets will also
provide a significant source of energy, and much detailed work has
been performed to follow up this suggestion. Strong evidence has
accumulated over recent years that the radiative losses of hot gas in
the 100\,kpc-scale cores of galaxy clusters may be counteracted by the
mechanical input from the central giant elliptical galaxies, which
usually host a low luminosity AGN (e.g. \citealt{csf+02}). It is
however not obvious that all AGNs, and in particular luminous quasars,
produce radio jets, whereas all do appear to produce high energy
radiation. In any case, the two mechanisms are complementary, and in
this paper we are exploring only the radiative feedback. Given what we
now know about the prevalence of MBHs in ellipticals, the tight
relationship between the MBH mass and stellar mass, and the known
efficiency for converting accreted mass into electromagnetic output
with a specific spectrum, the consequence of radiative feedback can
now be estimated with reasonable accuracy, more easily than can be
acheived for the mechanical energy feedback.

This paper is organized as follows. In Section~\ref{heatnear} we
investigate the possibility of radiatively preheating a spherical
accretion flow within 1--100\,pc from a MBH. In Section~\ref{heatfar}
we assess the conditions for a significant heating of interstellar gas
at larger distances from the central quasar. In Section~\ref{origin}
we propose that radiative feedback could be a key factor leading to
the observed $\mbh$--$\sigma$ relation. In Section~\ref{toy} we
elaborate on this idea in the context of the MBH--galaxy co-evolution,
using a simple, but physically motivated, one-zone model.

\section{Energetics of the radiative feedback}
\label{energy}

Radiative feedback from accretion onto a MBH can affect the accretion
flow within the 
MBH sphere of influence, i.e. in the central 1--100\,pc of the host
galaxy [see equations~(\ref{rbh}) and (\ref{sigma_mbh})], as well as the
interstellar gas at larger distances. Adopting the average quasar SED
from SOS, we consider the effects arising in the former and later case
in Section~\ref{heatnear} and \ref{heatfar}, respectively. We note
that significant feedback effects are expected only when the luminosity $L$
of the central MBH approaches the Eddington luminosity 
\beq
\Ledd=1.3\times 10^{46}\frac{\mbh}{10^8\ms}\,{\rm erg}\,{\rm s}^{-1},
\label{ledd}
\eeq 
hence our use of a SED typical of quasars throughout.

\subsection{Preheating of a spherical accreting flow onto a MBH}
\label{heatnear}

Using hydrodynamical simulations and fiducial quasar SEDs, CO01
demonstrated that accretion of gas onto the central MBHs in elliptical
galaxies cannot proceed stably at sub-Eddington rates if the
mass-to-radiation conversion efficiency $\epsilon$ ($\equiv\Delta
E_\gamma/\Delta mc^2$) is high ($0.1$), as is characteristic of
quasars. During high luminosity episodes, the high energy radiation,
primarily X-rays, from the MBH unbinds the accreting gas through
Compton heating, which temporarily leads to a decrease in the mass
inflow rate and consequently in the central luminosity. The SEDs
utilized by CO01 covered a fair range, in general corresponding to a
high Compton temperature. Our aim here is to verify whether the
conclusions remain unaltered for the average quasar SED once
photoionization is fully taken into account. We note that our
treatment below is similar to previous semi-analytic studies of
preheating of quasi-spherical accreting gas flows
\citep[e.g.][]{cos78,kl83,po99}.

\begin{figure}
\centering
\includegraphics[width=\columnwidth]{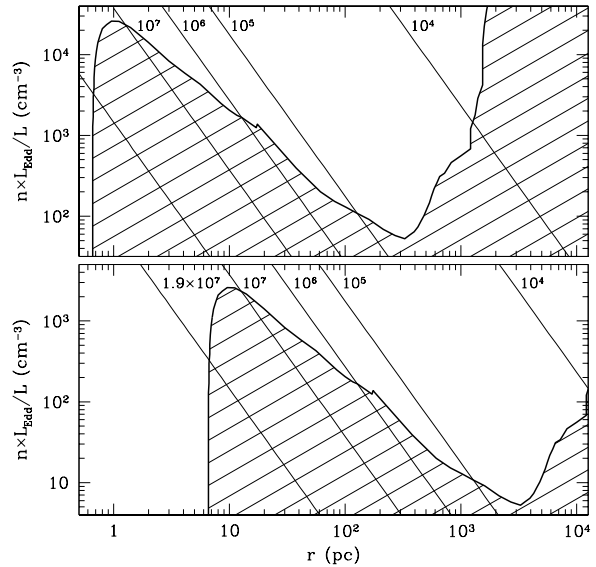}
\caption{Energy-based gas escape condition near a black hole of mass
$\mbh$ in the ($r$, $n$) plane, under the assumption that the gas is
in ionization equilibrium. Slanted lines correspond to constant
values of the ionization parameter as given by equation (3) for different
equilibrium gas temperature $\teq$ (labels along the lines). $\teq$ is
shown up to the Compton temperature ($1.9\times 10^7$\,K)
characterizing the average quasar.  The dashed area below the solid
curve given in equation~(\ref{mbh_blow}) corresponds to gas configurations
energetically unbound. In the upper and lower panel $\mbh=10^8\ms$ and
$\mbh=10^9\ms$, 
respectively.}
\label{nr_teq}
\end{figure}

SOS (see their fig.~5) computed for the average quasar SED the
equilibrium temperature $\teq$ of gas of cosmic chemical composition
as a function of the ``ionization'' parameter
\beq
\xi\equiv \frac{L}{nr^2},
\label{xi}
\eeq
where $n$ is the hydrogen number density and $r$ is the distance from
the MBH (note the use of the bolometric luminosity in the definition
of $\xi$). In the temperature range $2\times 10^4$--$10^7$\,K, to a
good accuracy
\beq
\teq(\xi)\approx 2\times 10^2\xi\,{\rm K},
\label{teq_xi}
\eeq 
while for $\xi\ll 100$ and $\xi\gg 5\times 10^4$, $\teq\approx
10^4$\,K and $2\times 10^7$\,K, respectively. $\teq$ is the
temperature at which heating through Compton scattering and
photoionization balances Compton cooling and cooling due to continuum and line
emission, on the assumption that the plasma is in ionization
equilibrium, appropriate for the problem under consideration. In
Fig.~\ref{nr_teq} we plot lines of constant $\teq$ on the ($r$, $n\ledd/L$)
plane for $\mbh=10^8\ms$ and $\mbh=10^9\ms$ [note that the full
expression of $\teq(\xi)$ is used for this purpose]. 

Suppose now that the gravitational potential is due to the central
MBH alone. Then the condition  
\beq
\frac{5}{2}k\teq(\xi)-\frac{G\mbh\mu\mpr}{r}>0
\label{mbh_blow}
\eeq
(here $\mu=0.61$ is the mean molecular weight) defines a situation
where gas of density $n$, located at $r$, will be heated to $\teq$ by the
central radiation and blown out of the MBH potential \footnote{We
note that in the case of the MBH emitting at near $\ledd$, the
radiation pressure will effectively reduce the MBH gravitational
potential, making possible an outflow in the directions not screened by the
central accretion disk for lower values of $\xi$ than implied by
equation~(\ref{mbh_blow}).}. For given $\mbh$, $L/\ledd$ and $r$, gas
with density below a certain critical value, as implicitly given by
equation~(\ref{mbh_blow}), cannot accrete onto the MBH, which
corresponds to the dashed areas in Fig.~\ref{nr_teq}.

We now specify the general conditions above to Bondi accretion. Note
that we are not assuming Bondi accretion onto the central MBH, since
the infalling material is expected to possess non-negligible angular
momentum and form an accretion disk or a more complicated central flow
near the MBH\footnote{In the case of quasars, it is believed that
accretion occurs via a geometrically thin, optically thick disk, the
presence of which naturally explains both the fact that most of the
radiative output takes place at ultraviolet wavelengths and the
inferred high radiative efficiency ($\epsilon\approx 0.1$) of
accretion (e.g. \citealt{yt02}). We briefly discuss the consequences
of the presence of a central accretion disk at the end of this
Section.\label{foot1}}.  We thus regard the Bondi relations as an
indication of the flow rates into the central regions as a source of
mass for the central flows.

The Bondi flow is transonic inside the radius   
\beq
\rbon=\frac{G\mbh\mu\mpr}{2\gamma kT}=16\,{\rm pc}
\frac{1}{\gamma}\frac{\mbh}{10^8\ms}\left(\frac{T}{10^6\,{\rm K}}\right)^{-1},
\label{rb}
\eeq 
where $T$ is the temperature at $\rbon$ and the sound speed
$\csound=\sqrt{\gamma kT/\mu\mpr}$. We shall assume $\gamma=1$
(isothermal flow) for estimations below, the results depending very
weakly on $\gamma$.  Comparison of equation~(\ref{mbh_blow}) with
equation~(\ref{rb}) shows that Bondi accretion of gas at temperature $T$
can be disrupted if the central luminosity $L$ is sufficiently high
that $\teq(\rbon)\gtrsim T$, or equivalently $L>\lcrit$, where
\beq
\lcrit(T)=\xi(T)\rbon^2(T) n(\rbon),
\label{lcrit}
\eeq
and $\xi(T)$ is the ionization parameter corresponding to
$\teq=T$. {\sl If $L>\lcrit$, the gas in the vicinity of $\rbon$ will
experience net radiative heating}. On the other hand, the central
luminosity is determined by the pre-existing conditions at $\rbon$ (so
far as $L<\ledd$):
\beq
L = \epsilon c^2\times 4\pi\rbon^2(T)\csound (T)\times\mu\mpr n_{\rm t}(\rbon),
\label{lb}
\eeq
where $n_{\rm t}\approx 2.3 n$ is the total particle number density.  

\begin{figure}
\centering
\includegraphics[width=\columnwidth]{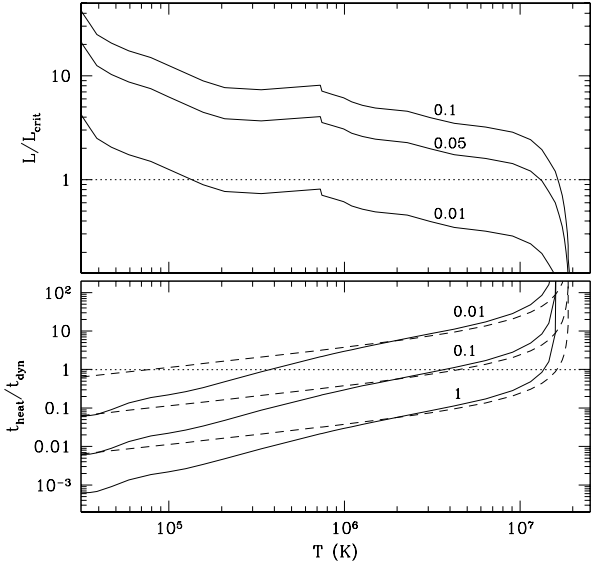}
\caption{Top panel: the ratio $L/\Lcrit$ as a function of $T$ for
Bondi accretion, as given by equations~(\ref{lcrit}) and (\ref{lb}). The
label above solid curves represents the adopted radiative
efficiency. $L/\Lcrit>1$ implies that stationary accretion is
impossible due to strong preheating. Bottom 
panel: ratio of the characteristic heating and flow times at the Bondi
radius, as given by equations (\ref{tflow}) and (\ref{theat}), as a
function of the gas temperature $T$ for $\epsilon=0.1$ and different
values of $L/\Ledd$ (labels). Solid lines correspond to net heating,
while dashed lines to Compton heating only. $\theat/\tflow>1$
implies that radiative heating is inefficient.} 
\label{preheat_limit}
\end{figure}

Fig.~\ref{preheat_limit} (upper panel) shows the ratio $L/\lcrit$ 
as a function of gas temperature for three values of the
accretion efficiency, $\epsilon=0.01$, 0.05 and 0.1. The two higher
values would correspond to accretion via a standard thin disk
\citep{ss73}. The presented curves have been obtained from
equations~(\ref{lcrit}) and (\ref{lb}) using the full expression of
$\teq(\xi)$, and can be approximated in a limited temperature range by the
expression
\beq
\frac{L}{\lcrit}\approx 6\frac{\epsilon}{0.1}
\left(\frac{T}{10^6\,{\rm K}}\right)^{-1/2},\,2\times 10^4\,{\rm
K}<T<10^7\,{\rm K},
\label{l_lcrit}
\eeq 
which follows from equation~(\ref{teq_xi}). Note that the ratio $L/\Lcrit$
does not depend on $\mbh$. The fact that $L>\Lcrit$ over most of the plot in
Fig.~\ref{preheat_limit} suggests that stationary accretion at high
rates is impossible, unless the gas supplied into the Bondi sphere
from outside is hotter than the Compton temperature of the average
quasar SED, i.e. $T>\tc\approx 2\times 10^7$\,K. If $T<\tc$, the
inflowing gas near $\rbon$ will be 
heated and driven out by the quasar radiation, and the MBH luminosity
will be reduced for the time required for significant amounts of gas
to cool and accumulate again in the central regions, i.e. the
result will be relaxation oscillations.

However, such a dramatic heating effect as described above will only
occur if the characteristic flow time at $\rbon$,
\beq
\tflow\equiv\frac{\rbon}{(2G\mbh/\rbon)^{1/2}}\approx
7\times 10^4\,{\rm yr}
\frac{\mbh}{10^8\ms}\left(\frac{T}{10^6\,{\rm K}}\right)^{-3/2},
\label{tflow}
\eeq
is longer than the heating time scale
\beq
\theat\equiv\frac{n_{\rm t}(\rbon)kT}{\Gamma(T,\xi)}.
\label{theat}
\eeq
Here $\Gamma(T,\xi)$ is the plasma net heating (cooling) rate per unit
volume, which in the case of the average quasar SED can be estimated
from formulae given in Appendix~\ref{app_agn}. Assuming
$\epsilon=0.1$, we find that for $L/\lcrit$ approximated by equation~(\ref{l_lcrit})
\[
\Gamma\approx 3\times 10^{-23}n^2(\rbon)\,{\rm erg}
\,{\rm s}^{-1}\,{\rm cm}^{-3},
\]
\[
\theat\approx 2\times 10^3\,{\rm yr}
\left(\frac{T}{10^6\,{\rm K}}\right)^{-1/2}\frac{\mbh}{10^8\ms}
\left(\frac{L}{\ledd}\right)^{-1},
\]
\beq
\frac{\theat}{\tflow}\approx 0.03\left(\frac{L}{\ledd}\right)^{-1}
\frac{T}{10^6\,{\rm K}},\,2\times 10^4\,{\rm K}<T<10^7\,{\rm K}.
\eeq
A more accurate computation leads to the curves shown in
Fig.~\ref{preheat_limit} (lower panel), giving the ratio $\theat/\tflow$
as a function of $T$ for different values of $L/\ledd$. Also plotted
is the ratio $t_{\rm C}/\tflow$ for the case of pure Compton
heating/cooling. One can see that Compton preheating is important when
$T\gtrsim 10^6$\,K and $L\gtrsim 0.1\ledd$. For $T<10^6$\,K, preheating due to
photoionization is significant at even lower accretion rates.

The above discussion suggests that if quasars are fed by warm ($T<
10^7$\,K) interstellar gas from distances $\sim$10--100\,pc, their
emission is expected to be significantly variable on time scales
$\sim\tflow\sim$$10^4$--$10^5$\,years. Only for gas temperatures
approaching $10^7$\,K, is stationary accretion at sub-Eddington rates 
possible (see Fig.~\ref{preheat_limit}).

We note however that, if the final stages of accretion
occur via a geometrically thin or slim disk, as expected for quasars (see
Footnote~\ref{foot1}), the above conclusion should change as regards
the variability time scale. Indeed, the characteristic viscous time in a
geometrically thin, optically thick accretion disk is given by \citep{ss73} 
\beqa 
t_{\rm visc}(r)\sim
3\times 10^5\,{\rm yr}\left(\frac{\alpha}{0.1}\right)^{-1}
\left(\frac{L}{\ledd}\right)^{-2}
\left(\frac{M}{10^8\ms}\right)^{-2.5} \nonumber\\
\times\left(\frac{r}{0.01\,{\rm pc}}\right)^{3.5},  
\label{tvisc}
\eeqa 
where $\alpha$ is the viscosity parameter. The above formula is valid
in the innermost, radiation pressure dominated zone of the disk, where
the main source of opacity is Thomson scattering.  
Such disks, characterized by a radially constant accretion
rate, are unlikely to extend beyond $10^{-3}$--$10^{-2}$\,pc because
they would become self-gravitating \citep{t64,ks80,g03}. Since the
quasar life time is expected to exceed $10^7$\,yr, the accretion disk
must be constantly replenished with gas of small angular momentum
(presumably from a Bondi-type flow). 

In the case of quasars ($\mbh\gtrsim 10^8\ms$, $L\gtrsim 0.1\ledd$),
self-gravity is expected to truncate the disk already in the innermost
zone, at the distance
\beq
R_{\rm sg}\approx 0.014\,{\rm pc}\left(\frac{\alpha}{0.1}\right)^{2/9}
\left(\frac{L}{\ledd}\right)^{4/9}
\left(\frac{M}{10^8\ms}\right)^{7/9},  
\eeq 
and the disk is expected to be stable against the thermal-viscous
instability \citep{bks98}. Comparison of equations~(\ref{tflow}) and
(\ref{tvisc}) implies that the timescale on which accreting matter
drifts from the outer boundary of the disk to the MBH, $t_{\rm
visc}(R_{\rm sg})$, can exceed the characteristic dynamical time of
the external Bondi-type flow, $\tflow(\rbon)$. In such a case,
variations in the gas inflow rate at $\rbon$ as a result of preheating
will affect the MBH accretion rate with a time delay $\sim t_{\rm
visc}$, leading to variations in the quasar luminosity also on time
scales $\sim t_{\rm visc}$.

\subsection{Heating of ISM in spheroids}
\label{heatfar}

The gravitational potential of the central MBH is overcome by the
potential of the host spheroid at distances larger than the MBH radius
of influence
\beq
\rbh\simeq\frac{G\mbh}{\sigma^2}\approx 10\,{\rm pc}
            \frac{\mbh}{10^8\ms} 
            \left(\frac{\sigma}{200\,{\rm km}\,{\rm s}^{-1}}\right)^{-2},
\label{rbh}
\eeq
where $\sigma$ is the characteristic (virial) one-dimensional stellar
velocity dispersion in the galaxy. Comparison of equation~(\ref{rbh}) with
equation~(\ref{rb}) implies that $\rbon<\rbh$ when $T\gtrsim\tvir$, where
\beq \tvir\simeq\frac{\mu\mpr\sigma^2}{k}=3.0\times 10^6\,{\rm K}
\left(\frac{\sigma}{200~{\rm km}~{\rm s}^{-1}}\right)^2
\label{tvir}
\eeq
is the galaxy virial temperature. Therefore, our discussion of
preheating in \S\ref{heatnear} pertains to situations where the
accreting flow has a temperature comparable to or higher than the
galaxy virial temperature.

Below we assess the conditions required for the central MBH to
significantly heat the interstellar gas over a substantial volume of
the galaxy, regardless of the type of accretion flow established in
the central regions. In this Section we shall {\sl assume} that the
MBH has a mass as given by the observed $\mbh$-$\sigma$ relation for
local ellipticals and bulges \citep{tgb+02}:
\beq
\mbh=1.5\times 10^8\ms\left(\frac{\sigma}{200~{\rm km}~{\rm
s}^{-1}}\right)^4. 
\label{sigma_mbh}
\eeq
Note that this assumption will be dropped in \S\ref{origin}, where we
attempt to {\sl predict} the $\mbh$-$\sigma$ relation.

\begin{figure}
\centering
\includegraphics[width=\columnwidth]{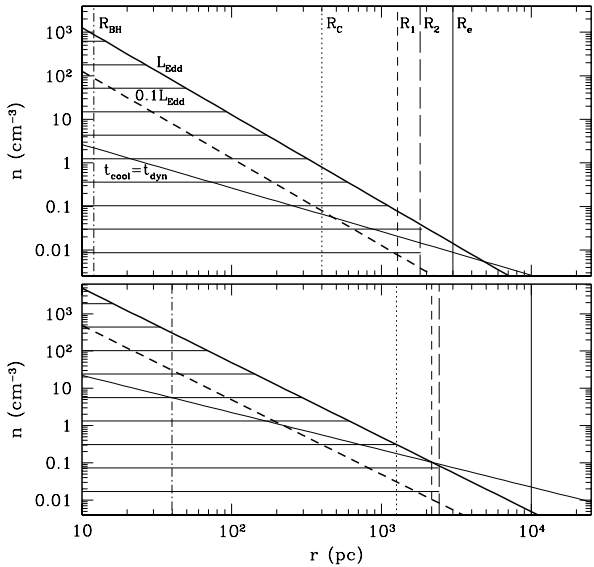}
\caption{The $(r,n)$ plane for a galaxy with $\sigma=180$\,km\,s$^{-1}$
($\Tvir=2.4\times 10^6$\,K, $\mbh=10^8\ms$, upper panel), and with
$\sigma=320$\,km\,s$^{-1}$ ($\Tvir=7.7\times 10^6$\,K, $\mbh=10^9\ms$,
lower panel). In the dashed area gas can be heated above $\Tvir$ by
radiation from the central MBH emitting at the Eddington
luminosity. The upper boundary of this area scales linearly with
luminosity (as shown by the dashed line corresponding to
$L=0.1\ledd$). From left to right vertical lines correspond to $\rbh$
[equation~(\ref{rbh})], $\rc$ [equation~(\ref{rc})], $R_1$
[equation~(\ref{r1})], $R_2$ [equation~(\ref{r2})], and $\reff$
[equation~(\ref{reff})]. The thin slanted line bounds from the above
the zone where the intrinsic cooling time of gas at $\tvir$ is longer
than the dynamical time. See text for further explanation.
}
\label{nr_tvir}
\end{figure}

From equations~(\ref{tvir}) and (\ref{sigma_mbh}) we can find the critical
density $\ncrit$, defined by
\beq
\teq(L/\ncrit r^2)=\Tvir,
\label{ncrit}
\eeq 
as a function of distance $r$ from the galaxy center. Gas with
$n<\ncrit$ will be heated above $\tvir$ and expelled from the galaxy,
whereas gas cooling will dominate over Compton and photoionization
heating if $n>\ncrit$. We show in Fig.~\ref{nr_tvir} the resulting
heating diagrams on the $(r,n)$ plane for two galaxies, with
$\sigma=180$ and 320\,km\,s$^{-1}$, corresponding to $\mbh=10^8$ and
$10^9\ms$, respectively. Also shown in Fig.~\ref{nr_tvir} is the
$n(r)$ line at which the cooling time $t_{{\rm cool},0}$ of gas at
$\tvir$ in the absence of radiative heating equals the characteristic
dynamical time, $\tflow\equiv r/\sigma$. Gas with density above this
line will cool down before reaching the galactic nucleus if there are
no heating mechanisms other than gravitational compression. Comparison
of the $\teq=\tvir$ line with the $t_{{\rm cool},0}=\tflow$ one
implies that if the gas within a few kpc of the MBH is sufficiently
tenuous to sustain a subsonic cooling flow (so that $t_{{\rm
cool},0}>\tflow$ and $T\approx\tvir$), the central quasar emitting
$L\gtrsim 0.1\Ledd$ will be able to heat the gas above $\tvir$. 

In practice, provided that $\teq>\tvir$, significant heating will take
place only out to a certain distance that depends on the luminosity
and duration of the quasar outburst. Since the MBH releases via
accretion a finite total amount of energy, $\epsilon\mbh c^2$, there
is a characteristic limiting distance given by:
\begin{eqnarray}
\rc &=&\left(\frac{\sigmat\epsilon\mbh}{3\pi\me}\right)^{1/2}
     = 400\,{\rm pc}\left(\frac{\epsilon}{0.1}\right)^{1/2}
          \left(\frac{\mbh}{10^8\ms}\right)^{1/2}\nonumber\\
    &=&500\,{\rm pc}\left(\frac{\epsilon}{0.1}\right)^{1/2}
          \left(\frac{\sigma}{200~{\rm km}~{\rm s}^{-1}}\right)^2. 
\label{rc}
\end{eqnarray}
Inside this radius, each electron--proton pair will have received
at least $3k\tc\approx 6$\,keV of energy through Compton scattering of
hard X-rays from the MBH when it has accreted mass $\mbh$. $\rc$
is defined assuming that the only heating mechanism is Compton
scattering, and thus pertains to the limit of low density gas, fully
photoionized by the radiation from the MBH. 

More relevant for the problem at hand is the distance out to
which low density gas will be Compton heated to $T\ge\Tvir$:
\beq
R_1=\rc\left(\frac{\tc}{\tvir}\right)^{1/2}=
    1,300\,{\rm pc}\left(\frac{\epsilon}{0.1}\right)^{1/2}
       \frac{\sigma}{200~{\rm km}~{\rm s}^{-1}}.
\label{r1}
\eeq\
Yet another characteristic radius is the one within which gas of critical
density $\ncrit$ will be heated to $T\ge\Tvir$ by photoinization
and Compton scattering:   
\beq R_2=R_1\left[\frac{\Gamma(n_{\rm crit})}{\Gamma_{\rm C}}\right]^{1/2},
\label{r2}
\eeq
where $\Gamma_{\rm C}$ and $\Gamma$ are the Compton and total heating rates,
respectively. Depending on the gas density ($0<n<n_{\rm crit}$), the
outer boundary of the ``blowout region'' will be located somewhere
between $R_1$ and $R_2$. The size of the heating zone may be compared (see
Fig.~\ref{nr_tvir}) with the galaxy effective (half-light) radius
$\reff$. We estimate from the results of \citet{f+97} and \citet{bsa+03}
that for early-type galaxies  
\beq
\reff\sim 4,000\,{\rm pc}
\left(\frac{\sigma}{200~{\rm km}~{\rm s}^{-1}}\right)^2,
\label{reff}
\eeq
and note that in reality the correlation between $\reff$ and $\sigma$
is fairly loose.

\begin{figure}
\centering
\includegraphics[width=\columnwidth]{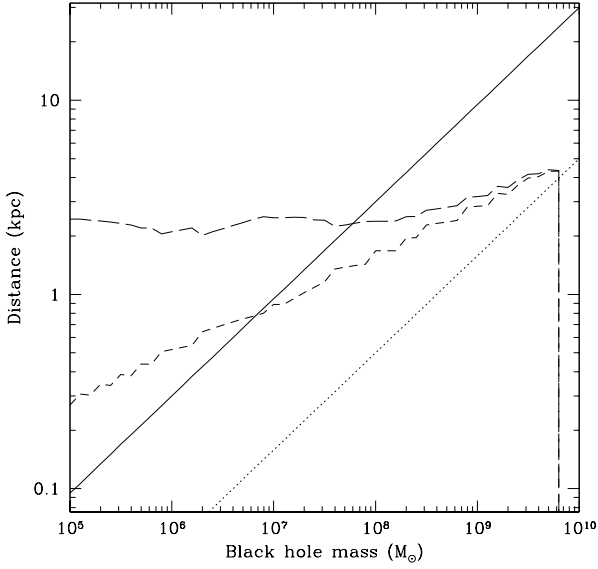}
\caption{Different characteristic heating radii, $\rc$ (dotted line),
$R_1$ (short-dashed line), and $R_2$ (long-dashed line), as a
function of $\mbh$, compared with the general trend for the effective
radius of elliptical galaxies (solid line).
}
\label{mbh_radii}
\end{figure}


The different characteristic distances defined above are shown as a
function of $\mbh$ in Fig.~\ref{mbh_radii}. One can see that the
radiative output of a black hole of mass $<10^7\ms$ can unbind the
interstellar gas out to several $\reff$ in relatively low luminosity
elliptical galaxies. In the case of more massive black holes/galaxies
with $\mbh\sim$$10^8$--$10^9\ms$, the heating will be localized to the
innermost 0.3--0.5$\reff$; the gas further out can be heated only
indirectly, presumably through shock waves propagating from the
radiatively heated central regions (see CO01). Remarkably,
$R_2\approx$\,2--4\,kpc, almost independently of $\mbh$. This results
from a combination of two opposite trends: the lighter the MBH is, the
less total energy it emits, and the lower $\tvir$ is, the more
important becomes the role of photoionization heating compared to
Compton heating for gas with $n\lesssim\ncrit$. We point out that
radiation characterized by the average quasar SED could not unbind the
gas in galaxies with $\sigma>500$\,km\,s$^{-1}$ ($\mbh>6\times
10^9\ms$) if such giant galaxies existed, because then
$\tvir>\tc\approx 2\times 10^7$\,K. On the contrary, the gas in the
central regions of such galaxies would be {\sl Compton cooled} by the
radiation from the central quasar. Presumably it is a coincidence that
the upper limit to the velocity dispersion for real elliptical
galaxies is close to the Compton temperature found for the radiation
from typical massive quasars.

We finally note that in the case of a quasar outburst with given
$L$ and duration $t$, during which a relatively small amount of mass
$\Delta\mbh\ll\mbh$ is accreted, the radiative heating front will
propagate out to 
\beq
\tilde{R}_{1,2}(L,t)=R_{1,2}\left(\frac{L}{\ledd}\right)^{1/2}
\left(\frac{t}{2\times 10^7\,{\rm yr}}\right)^{1/2}
\left(\frac{\epsilon}{0.1}\right)^{-1/2}
\eeq
in time $t$, where $R_{1,2}$ are given by equations~(\ref{r1}),
(\ref{r2}). It is worth noting that in the case of Eddington-limited
accretion, the characteristic heating time at $R_{1,2}$ is $2\times
10^7(0.1/\epsilon)$\,yr, or approximately half the Salpeter time scale
(time required for the black hole mass to double).


\section{The proposed origin of the $\mbh$--$\sigma$ relation}
\label{origin}

We now address the central issue of this work, namely the possibility
that radiative feedback  played the key role in establishing the
observed $\mbh$--$\sigma$ relation. 

Below we elaborate on the following general idea. Before the MBH grows
to a certain {\sl critical mass}, $\mcrit$, its radiation will be
unable to efficiently heat the ambient gas, and accretion onto the MBH
will proceed at a high rate. Once the MBH has grown to $\mcrit$, its
radiation will heat and expel a substantial amount of gas from the
central regions of 
the galaxy\footnote{Quite obviously, this effect will cooperate
with the energy input from stellar winds and SNII explosions in the
forming galaxies, however here we show that in principle the radiative
feedback from the MBH only is already sufficient to do the
work.}. Feeding of the MBH will then become self-regulated on the
cooling time scale of the low density gas. Subsequent quasar activity
will be characterized by a very small duty cycle ($\sim$0.001), as
predicted by hydrodynamical simulations (CO97, 01) and suggested by
observations \citep{hco04,hkb+04}. MBH growth will be essentially
terminated.

Suppose that the galaxy density distribution is that of a
singular isothermal sphere, with the gas density following the total
density profile:
\beq 
\rho_{\rm gas}(r)=\frac{\mgas}{M}\frac{\sigma^2}{2\pi Gr^2}.
\label{rho_gas}
\eeq
Here $\mgas$ and $M$ are the gas mass and and total mass within the
region affected by radiative heating. The size of the latter is
uncertain but is less than a few kpc (see \S\ref{heatfar}), so that
$M$ is expected to be dominated by stars ($\mstar\lesssim M$) rather than by
dark matter.

Radiation from the central MBH can heat the ambient gas up to the
temperature 
\beq
\teq\approx 6.5\times 10^3\,{\rm K}
\frac{L}{\ledd}\left(\frac{\mgas}{M}\right)^{-1}\frac{\mbh}{10^8\ms}
               \left(\frac{200\,{\rm km~s}^{-1}}{\sigma}\right)^2.
\label{tstat}
\eeq
This approximate relation is valid in the range $2\times
10^4$--$10^7$\,K and follows from equations~(\ref{teq_xi}) and
(\ref{rho_gas}). Remarkably, $\teq$ does not depend on distance for
the adopted $r^{-2}$ density distribution. We then associate the
transition from rapid MBH growth to slow, feedback limited MBH growth
with meeting the critical condition  
\beq
\teq=\etaesc\tvir, 
\label{tstat_tvir}
\eeq
where $\etaesc\gtrsim 1$ and $\tvir$ is given by equation~(\ref{tvir}). Once
heated to $\teq=\etaesc\tvir$, the gas will stop feeding the MBH. We
readily find that the condition (\ref{tstat_tvir}) will be met for
\beq
\mcrit=4.6\times 10^{10}\ms\etaesc\left(\frac{\sigma}
       {200\,{\rm km~s}^{-1}}\right)^4\frac{\ledd}{L}\frac{\mgas}{M}.
\label{mcrit}
\eeq 
Therefore, for fixed values of $\etaesc$, $L/\ledd$ and $\mgas/M$ we
expect $\mcrit\propto\sigma^4$, similar to the observed
$\mbh$--$\sigma$ relation. According to our proposed scenario, once
the MBH has reached the critical mass, its accretion growth will be
effectively terminated so that at the present epoch $\mbh$ is expected
to be only slightly larger than $\mcrit$. 

Equally important information is contained in the
normalization of the $\mbh$--$\sigma$ relation. In fact, by comparing
equation~(\ref{mcrit}) with equation~(\ref{sigma_mbh}) we find that
the observed relationship will be established if 
\beq 
\frac{\mgas}{\mstar}= 3\times
10^{-3}\etaesc^{-1}\frac{L}{\ledd}\frac{M}{\mstar}.
\label{mgas_crit}
\eeq 
To satisfy the observed $\mbh$--$\sigma$ relation, the gas-to-stars
ratio is thus required to be relatively low and approximately constant
for spheroids with different masses at the epoch when the MBH reaches
its critical mass, although the observational uncertainty in the
$\mbh$--$\sigma$ relation leaves some room for a weak dependence of
$\mgas/\mstar$ on $\sigma$. As for the Eddington ratio, it is reasonable to
expect $L/\ledd\sim$0.1--1 based on hydrodynamical simulations (CO01)
and observations of quasars (see e.g. \citealt{hco04} and references therein).

\begin{figure}
\centering
\includegraphics[width=\columnwidth]{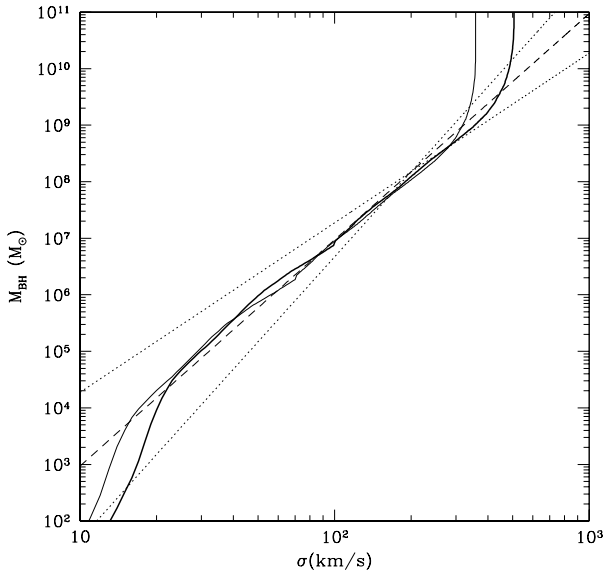}
\caption{Thick solid line shows the predicted $\mbh$--$\sigma$
relation resulting from the requirement that heating of the
interstellar gas by radiation from the central MBH at the
Eddington limit be below the level required to drive the gas from the
galaxy ($\teq\leq\tvir$). This upper bound on $\mbh$ is based on
assuming a constant gas fraction of $\mgas/M=0.003$ and
$\etaesc=1$. The thin solid line corresponds to $\mgas/M=0.0015$ and
$\etaesc=2$. The dashed line is the observed $\mbh\propto\sigma^4$
relation in the range $\mbh=10^6$--a few $10^9\ms$, extrapolated to
lower and higher $\mbh$ values from equation~(\ref{sigma_mbh}). The
dotted lines are $\mbh\propto\sigma^3$ and $\mbh\propto\sigma^5$ laws.}
\label{s_mbh}
\end{figure}

We note that the above calculation was based on a specific ($r^{-2}$)
gas density profile for which  $\xi=L/nr^2=const$. This allowed us to
avoid specifying the size of the radiative heating zone. Being a
reasonable assumption for the outer regions of the galaxy, the
$r^{-2}$ profile is however expected to flatten in the central
$\sim$$0.1\reff$ region (see, e.g., CO01), so that $\xi$ will be
increasing toward the MBH. Since it can be sufficient to overheat the
cooling gas in this central zone to produce dramatic effects on the
subsequent co-evolution of the galaxy and its central MBH, our scenario
certainly allows the critical $\mgas/\mstar$ ratio (defined say within
$\reff$) to be several times larger than required by
equation~(\ref{mgas_crit}).

We note that the approximately linear $\teq(\xi)$ dependence [see
equation~(\ref{teq_xi})] was crucial to the above argument leading to the
$\mcrit\propto\sigma^4$ result. However, the $\teq(\xi)$ function
becomes strongly nonlinear outside the range $2\times 10^4\,{\rm
K}<\teq<10^7\,{\rm K}$, and a more general result can be obtained if
we consider the exact curve $\teq(\xi)$ from SOS. In Fig.~\ref{s_mbh}
we show the predicted correlation between $\mcrit$ and $\sigma$ for
$\etaesc=1$, $L/\ledd=1$ and $\mgas/M=3\times 10^{-3}$, and compare it
with the observed relationship. We see that the $\mbh\propto\sigma^4$
behavior is expected to break down for $\mbh<10^4\ms$ and also for
$\mbh\gtrsim$\,a few\,$10^9\ms$. In the same figure we demonstrate the
effect of the escape parameter $\etaesc$. For $\etaesc=2$ and the gas
fraction decreased two-fold, the $\mbh$--$\sigma$ relation is
unchanged except that the high-mass cutoff occurs at a lower
mass. Specifically $M_{\rm BH, cut}\approx 3\times
10^9\ms\etaesc^{-2}$.

It is perhaps interesting that the range of masses shown in
Fig.~\ref{s_mbh} for which $\mbh\propto\sigma^4$ is obtained from
considerations of atomic physics (and the observed AGN spectra)
corresponds closely with the range of masses for which this power law
provides a good fit to the observations. Exploring the
$\mbh$--$\sigma$ relation observationally near $10^9\ms$ would be a
sensitive test of the importance of radiative feedback. We note that
other scenarios discussed in the literature also 
predict deviations from the power law trend for the most massive MBHs 
(e.g. \citealt{m+03}). 

\section{A simple, physically based toy-model}
\label{toy}

In this Section we address in a more quantitative way the MBH--galaxy
co-evolution. We adopt a physically-motivated one-zone model described
in detail in Appendix~\ref{app_toy}, focusing in particular on the
co-evolution of the galaxy gas budget and stellar mass, a key
ingredient of the proposed scenario (see Sect. 3). Several aspects of
this model have been already described in Ciotti, Ostriker \& Sazonov
(2004) and Ostriker \& Ciotti (2004).

In particular, source terms for the galaxy gas mass
(equation [\ref{dmgas}]) are due to cosmological infall (equation [\ref{dminf}])
and to stellar mass return from the evolving stellar population
(equation [\ref{dmrec}]), while gas is subtracted from the total budget by
star formation (equation [\ref{dmstar}]), gaseous MBH accretion
(equation [\ref{dmbhac}]), and (possible) galactic winds when the thermal
energy of the ISM is high enough to escape from the galaxy potential
well (equation [\ref{dmesc}]). Energy input on the galactic gas is due to
thermalization of SNII and SNIa explosions, to thermalization of red
giants winds due to the galaxy stellar velocity dispersion
(Sects. A3.1 and A3.2), and to the radiative feedback from the
accreting MBH (Sect. A3.3). In case of galactic winds, we also
consider adiabatic cooling of the expanding gas (Sect. A3.4). Stars
are formed by gas cooling, while {\it two} different mechanisms are
considered for the MBH growth, namely gaseous accretion and
coalescence of stellar remnants of massive stars (equation \ref{dmbh}). The
galaxy potential well is determined only by the dark matter potential
well, which is assumed to be unevolving with cosmic time. In other
words, here we are not considering merging between galaxies; finally,
we note that in the present toy-model the possibility to add a simple
recipe for chemical evolution is straightforward, so that one could
also test the model against scaling laws such as the
Mg${_2}$--$\sigma$ relation.

We remark here that this kind of approach is not new
\citep[e.g.][]{gds+04,m+03}. However, our scenario presents a few but
important differences with respect to the other cases. For example, in
their modelization of the AGN feedback \citet{gds+04} (see also
\citealt{mqt04}) assumed that the main role is played by radiation
pressure through scattering and absorption by dust and scattering in
resonance lines, while the energy release from the AGN to the ISM is
due to thermalization of the kinetic energy of the outflow. In our
case, instead, the feedback is due to {\it radiative} heating. Note
that energy absorption is typically more efficient in
driving winds than momentum absorption.

Obviously, the results of such an approach should not be
``overinterpreted'': as common in all similar approaches, the
parameter space of the present model is huge (even though several
input parameters are nicely constrained by theory and/or
observations), thus the results of such kind of simulations should be
interpreted more as indications of possible evolutionary histories
than exact predictions. In particular, the {\it toy-model cannot
directly test the ability of radiative feedback to produce the right
final MBH mass}, in fact, this can be done only using numerical
hydrodynamical simulations. This is not surprising, because the
toy-model, by construction, is a {\it one zone} model, and we already
know that feedback mechanisms are strongly scale-dependent, in the
sense that central galaxy regions react in a substantially different
way with respect to the whole system (CO97, CO01,
Sect.~\ref{heatfar}).

As a consequence, one of the key input parameters in our scheme is the
adopted ``quasar duty cycle'' $\fed$ [equation~(A6)], which essentially
introduces a limit, $\fed\ledd$, on the instantaneous central quasar
luminosity that is smaller than the Eddington luminosity. In the
following, we present the results for two different assumptions about
$\fed$. In the first (Sect.~\ref{reduced}), we distinguish two
evolutionary phases: a first phase (the ``cold phase'', in which
$\tcool/\tdyn <1$) that would be identified observationally with the
Lyman Break Galaxies, and a later phase (the ``hot phase'', in which
$\tcool/\tdyn >1$) that would be identified with normal, local
ellipticals which contain little gas, have low rates of star formation
and have a duty cycle (fraction of time during which they appear as
luminous AGNs) of roughly 0.1\%.  Thus, in equation~(A6) the duty cycle
factor is known empirically to be $\fed\simeq 0.01$ in the cold phase
(where roughly a few per cent of Lyman Break Galaxies show central
AGNs, \citealt{s+02,l+04}), while $\fed \simeq 0.001$ in the hot
phase.  It is important to note that the last value is suggested by
both numerical simulations (CO01) and observations \citep{hco04}.

A different model is also explored (Sect.~\ref{unity}), in which
$\fed$ is kept fixed to 1 over all the simulation. This case will
adequately demonstrate the effects of radiative feedback on the
toy-model evolution.

\subsection{Models with reduced $\fed$}
\label{reduced}

\begin{figure}
\centering \includegraphics[width=\columnwidth]{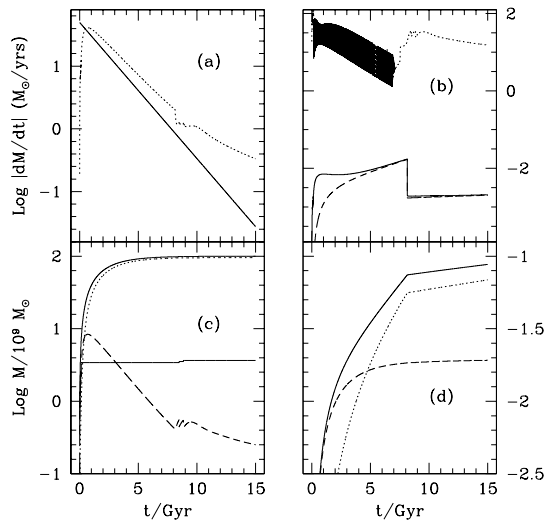}
\caption{Reference model. Panel $a$: mass infall rate (solid line) and 
stellar mass formation rate (dotted line). Panel $b$: total (gaseous
plus stellar remnants) MBH accretion rate (solid line), Bondi
accretion rate (dotted line), Eddington accretion rate (reduced by the
duty-cycle factor $\fed$, dashed line).  Panel $c$: total infall mass
(solid line), total stellar galaxy mass (dotted line), total galaxy
gas mass (short-dashed line); the nearly horizontal line is the
escaped gas mass. Panel $d$: total MBH mass (solid line), total mass
gaseously accreted (dotted line), MBH mass originated from stellar
remnants (dashed line).}
\label{mod1a}
\end{figure}

\begin{figure}
\centering\includegraphics[width=\columnwidth]{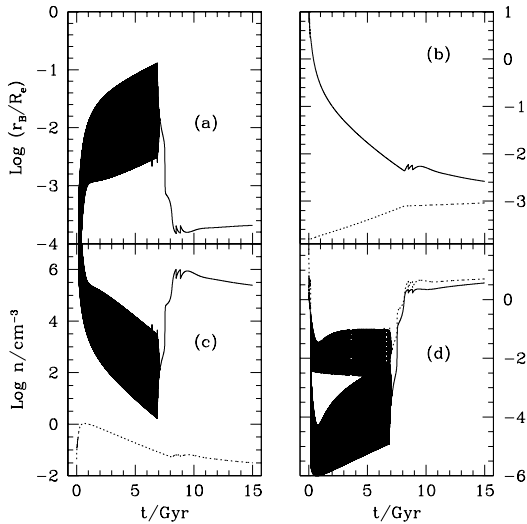}
\caption{Reference model. Panel $a$: time evolution of the Bondi
radius.  Panel $b$: logarithm (base 10) of the ratio between gas mass
to stellar mass (solid line) and of the ration between MBH mass to
stellar mass (``Magorrian relation'', dotted line).  Panel $c$: gas
density at the Bondi radius (solid line) and mean gas density (dotted
line). Panel $d$: logarithm (base 10) of the cooling time (solid line)
and heating time (dotted line) measured in terms of the dynamical tyme.}
\label{mod1b}
\end{figure}

\begin{figure}
\centering\includegraphics[width=\columnwidth]{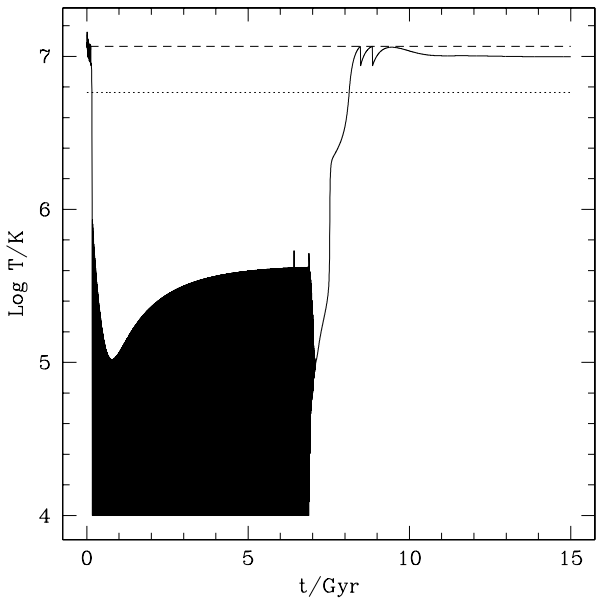}
\caption{Time evolution of the model gas temperature (solid line). The model 
virial temperature is represented by the dotted line, while the dashed line 
represent the ``escape'' temperature (here assumed $2\Tvir$).}
\label{mod1c}
\end{figure}

With the above remarks in mind, here we present a few representative
simulations primarily aimed at investigating whether there exists a
specific evolutionary phase characterized by a gas-to-star mass ratio
of the order of the factor in equation~(\ref{mgas_crit}), required to
obtain the right $\mbh$--$\sigma$ relation. The time evolution of the
quantities shown in Figs.~\ref{mod1a}--\ref{mod1c} refers to a galaxy
reference model characterized by $\Reff=4$\,kpc and a halo (constant)
circular velocity of 400\,km\,s$^{-1}$; the total mass of the gas
infall is $10^{11}\ms$, and the characteristic infall time is 2\,Gyr.
Other relevant simulations parameters are $\alstar =0.3$ in
equation (\ref{dmstar}), $\bhstar = 1.5\times 10^{-4}$ in equation (\ref{dmbh}),
$\epsilon=0.1$ in equation (\ref{Aedd}), $\etasn=0.5$ in equation (\ref{desn}),
and finally $\etaesc=2$ in equation (\ref{dmesc}). The initial black hole
mass is assumed to be $10\ms$, and the duty-cycle is fixed according
to the prescription of the ``cold/hot'' phases described above.


While a complete description of the toy-model behavior for different
choices of the input parameters is outside the scope of this paper,
here we remark that after an initial ``cold'' phase dominated by gas
infall, as soon as the gas density becomes sufficiently low
(Fig.~\ref{mod1a}c), and correspondingly the cooling time becomes
longer than the dynamical time (Fig.~\ref{mod1b}d), the gas heating
dominates, and the galaxy switches to a ``hot'' solution
(Fig.~\ref{mod1c}). The gas mass/stellar mass ratio at that moment
($\sim 0.003$, Fig.~\ref{mod1b}c) is very near with the value inferred
in Section~\ref{origin} from the argument leading to the right
$\mbh$--$\sigma$ relation. Note also how the gas content of the
present day ``galaxy'' is in nice agreement with observations.  It
proves that if $\fed$ is sufficiently small (as in the present case),
the accretion is Eddington (rather than Bondi) limited (equation [A6])
throughout the MBH--galaxy co-evolution, and then our model increases
the MBH mass in the prescribed way {\it independently} of the specific
feedback mechanism. The obtained final black hole mass is in nice
agreement with the Magorrian relation (dotted line, Fig. 7b), but the
significance of this result should not be overestimated since in
reality the Magorrian relation is set (in our scenario) by the MBH
feedback.

An interesting experiment is obtained by reducing the circular halo
velocity and the infall mass in the reference model: in these cases
galactic winds are favoured. In other words, small galaxies lose
their gas content easily, in accordance with the Mg$_2$--$\sigma$ and
the Faber--Jackson relations and with the hydrodynamical simulations of
CO01. Remarkably, the transition to the hot phase of these models
happens for $\mgas/\Mstar\sim 0.01$, similarly to the behavior seen in
the case of the more massive spheroid in the reference model. In other
models we have verified the role of the AGN photoionization by
artificially excluding the sources of stellar heating in the reference
model: in this case the galaxy gas temperature (after the initial cold
phase) remains sub-virial (i.e. below the horizontal dotted line in
Fig.~8), without however cooling down to the imposed lower temperature
limit of $10^4$\,K, as instead happens if we also 
exclude the MBH feedback. This sort of cooperation between AGN
feedback and stellar energy injection, i.e., the fact that {\it
substantial galactic winds in general are due to stellar heating, and
are reinforced by the presence of the central AGN}, was already found
in numerical simulations (CO01). 

An important and apparently robust conclusion that can be drawn from
these simulations is that stellar heating inevitably leads
to a transition from cold to hot solution when the gas-to-star mass
ratio drops to of order 1 per cent or somewhat less. Now, since a gas
fraction of this order is required for the radiative feedback from the
central MBH to limit its growth at the mass obeying the observed
$\mbh$--$\sigma$ relation (see Sect.~\ref{origin}), it is tempting to
suggest that the MBH reaches its critical mass, determined by
radiative feedback, approximately at the epoch of transition from cold
to hot galaxy phase.  

We emphasize again that in the above models the radiative output from
the MBH does not strongly influence the interstellar gas, since $\fed$ is 
assumed to be low and thus the central luminosity is forced to be less
than $\fed\ledd\ll\ledd$ at any instant. In other words, these models
allow persistent moderate AGN activity but forbid quasar-type outbursts. 

\subsection{A model illustrating radiative feedback}
\label{unity}

In order to illustrate the effects of MBH feedback in the
context of the toy model, we now present a model in all similar to
reference model (Sect.~\ref{reduced}), but in which $\fed =1$ over all
the evolution. In this model we also reduce the characteristic infall
time from 2\,Gyr to 0.2\,Gyr, and we set $\bhstar = 0$ in
equation (\ref{dmbh}), i.e., we are neglecting black hole growth due to
accretion of stellar remnants. 

Several differences are apparent with respect to the evolution of
the reference model and its variants. In fact, the final black hole mass is
now larger, the final amount of gas in the galaxy is significantly
lower, and the galaxy is found in a permanent wind state (excluding
the very short initial cold phase). Also, from Fig. 9b is apparent how
the MBH grows by Bondi accretion instead of being Eddington
limited (again excluding the initial phase). The main difference,
however, is that now a very strong gas outflow is caused by the MBH
feedback rather than by energy input from the evolving stellar
population. 

Most importantly, this model supports the argument presented in
Section~\ref{origin} in that quasar radiative heating terminates the MBH
growth at a mass proportional to the gas fraction at the critical
epoch, i.e. when the MBH feedback becomes important. The fact that in
this particular simulation the final black hole mass turns out to be
about 2 orders of magnitude higher than the ``Magorrian'' mass is due
to the fact that the radiation from the central quasar overheats the
gas when its mass fraction is $\sim 10$ per cent (see Fig.~10b), much higher
than a fraction of 1 per cent. This happens in this model because the
Eddington limited growth of the black hole allowed it to reach a very
high mass while the galaxy gas content has not yet been reduced to low
levels.

An obvious problem with the models described in
Sections~\ref{reduced} and \ref{unity} is that the parametrization of
the MBH accretion rate and luminosity in terms of the time averaged
$\fed$ factor is a very poor substitute for considering the real, time
dependent problem. Nevertheless, we believe that the models assuming a
small $\fed$ and the model in which $\fed=1$ nicely complement each other and
provide important clues to the real picture of MBH--galaxy
co-evolution. Specifically, the first kind of models demonstrate that
there is indeed a well defined phase in galactic evolution when the
gas-to-stars mass ratio is of the order of 1 per cent -- it corresponds
to the transition from cold to hot solution. On the other hand, the
$\fed=1$ model demonstrates how the MBH radiative feedback can be
efficient if the central quasar switches on at a near Eddington
luminosity. Taken together, these models suggest that a major quasar
outburst occuring approximately at the beginning of the hot galaxy
phase, when $\mgas/\Mstar\sim 0.01$ or somewhat less, can lead 
to a significant degassing of the galaxy and termination of the MBH
growth at the mass required by the $\mbh$--$\sigma$ relation.

\begin{figure}
\centering \includegraphics[width=\columnwidth]{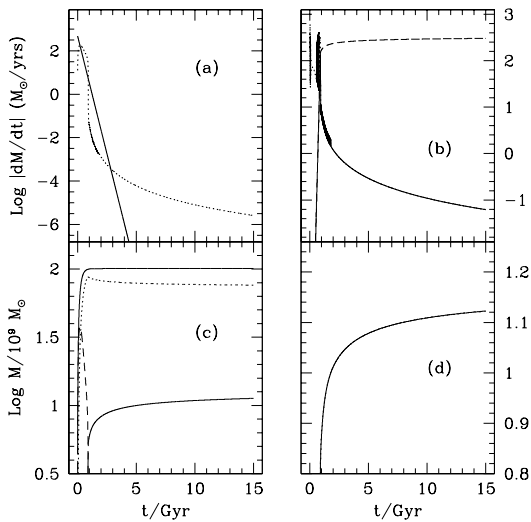}
\caption{Model with $\fed=1$ and $\tinf=0.2$ Gyr. See Caption of Fig. 6 for a 
description of the various curves. Note also how in the hot phase 
MBH accretion is Bondi dominated.}
\label{mod2a}
\end{figure}
\begin{figure}
\centering \includegraphics[width=\columnwidth]{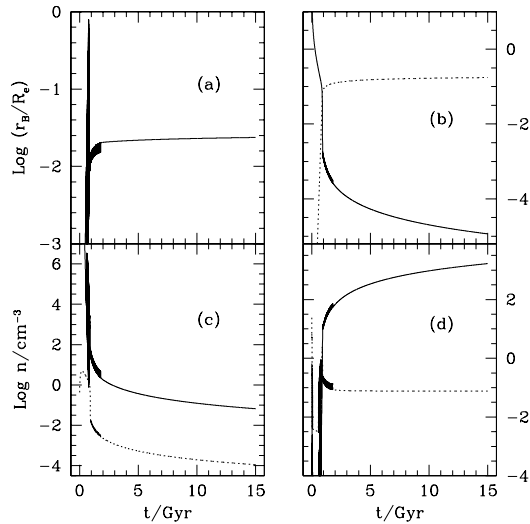}
\caption{Same model as in Fig. 9. See Caption of Fig. 7 for a 
description of teh various curves.}
\label{mod2b}
\end{figure}
\begin{figure}
\centering\includegraphics[width=\columnwidth]{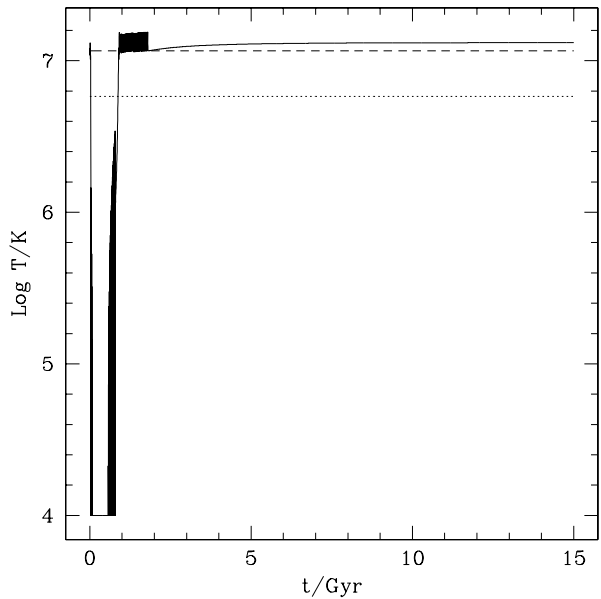}
\caption{Temperature evolution of the model in Figs. 9 and 10. 
See Caption of Fig.8 for a description of the various curves. Note how the 
galaxy is found in a permanent wind phase after $\simeq 1$ Gyr.}
\label{mod2c}
\end{figure}

\section{Discussion and conclusions}
\label{conclusions}

In this paper we explore the possible role of AGN radiative feedback in the
co-evolution of MBHs at the center of spheroids and their stellar and
gaseous components. The feedback is due to a combination of
photoionization and Compton heating. In our calculations we adopt
(from the work of SOS) a broad-band spectral energy distribution that
corresponds closely with the radiative output of typical quasars.

We first investigate (Sect.~\ref{heatnear}), based on energetic
considerations, whether the radiative output of the central MBH can
significantly affect a Bondi-type accretion flow in the central
1--100\,pc of a galaxy. In agreement with previous studies, we find
that if the accretion is radiatively efficient and proceeds at a subcritical
rate, as expected of quasars, the mass inflow rate and the MBH
luminosity are expected to oscillate on time scales $\sim
10^4$--$10^5$\,yr as a result of gas preheating, unless the
temperature of the supplied interstellar gas approaches the Compton 
temperature of the radiation field ($2\times 10^7$\,K). We also note
that if a central accretion disk extends beyond
$\sim$\,0.001--0.01\,pc from the MBH, then preheating of the external,
hot flow may still lead to variations in the MBH accretion rate, but
on longer time scales (up to millions of years), determined by the
characteristic drift time in the disk. 

Using similar energetic considerations, we then assess the effect of
the MBH radiative feedback on the main volume of the host galaxy 
(Sect.~\ref{heatfar}). We demonstrate that the radiative output from a
MBH growing by radiatively efficient accretion can unbind the ambient
interstellar gas, provided it is tenuous enough, out to a maximum
distance of a few kpc, which would typically amount to a significant
fraction of the effective radius for giant ellipticals and to a few 
effective radii for smaller spheroids.

We then discuss (Sect.~\ref{origin}) a possible origin of the observed
$\mbh$--$\sigma$ relation based on the hypothesis that the radiative
output from the growing MBH will eventually heat the ambient
interstellar gas above the virial temperature and expel most of it,
limiting both future growth of the MBH and future star formation to
low levels.  We demonstrate that if the gas-to-stars mass ratio drops
to $\lesssim 1$ per cent in the central regions of spheroids at a
certain stage of their evolution, then the radiative feedback from the
central MBH switched on as a bright quasar will terminate its growth
at the mass obeying the observed $\mbh$--$\sigma$
relation. Furthermore, we predict that the observed power law should
break down for black hole masses below $10^4\ms$ and above a few
$10^9\ms$. These considerations however leave open the question as to
why the gas fraction should be of the above order when the MBH reaches
the critical mass.

In order to obtain a better insight into the co-evolution of gas,
stars and the central MBH over the Hubble time we explore a simple but
physically based one-zone toy model (Sect.~\ref{toy}). A robust result
obtained from the simulations is that when the gas mass fraction has
been reduced by star formation to of order 1 per cent or somewhat
less, a transition from cold to hot solution takes place, both in the
case of massive protogalaxies that would evolve into the present day
giant ellipticals and in the case of smaller spheroids. This
transition occurs primarily as a result of gas heating due to the
evolving galaxy stellar component in its various forms (stellar winds
thermalization, SNII and SNIa explosions), while the AGN heating is
also expected to contribute at some level.

The near coincidence of the gas fraction corresponding to the
beginning of the hot galactic phase with that [equation~(\ref{mgas_crit})]
required by our argument leading to the correct $\mbh$--$\sigma$
relation offers the possibility of the following evolutionary
scenario. At the early stages of galaxy evolution when the
protogalactic gas is dense and cold, active star formation is
accompanied by the growth of a central black hole. The black hole
however is not massive enough to produce a strong heating effect on
the ambient, dense gas, even during episodes when it shines near at
the Eddington limit. This cold phase would be identified
observationally with the Lyman Break Galaxies and bright submillimeter
galaxies, which are characterized by high star formation rate and
moderate AGN activity. The cold phase ends when the gas-to-stars mass
ratio has been reduced to $\sim 0.01$, when the energy input from the
evolving stellar population and possibly from the central MBH heat the
gas to a sub-virial temperature. The MBH continues to grow actively
during this transitional epoch (perhaps more actively than before)
because there are still sufficient supplies of gas for accretion, and
soon reaches the critical mass (obeying the $\mbh$--$\sigma$
relation), when the MBH radiative output causes a major gas
outflow. This phase would be identified with the major quasar
epoch. The subsequent evolution is passive and characterized by very
low AGN activity and a duty cycle reduced by a factor of ten to
0.001, and this late phase would be identified with the present day
elliptical galaxies.

This scenario is admittedly tentative and not the only possible one,
and should be verified by observations and more detailed
computations. In particular, it should be elucidated as to why the MBH
does not grow to an excessive mass during the cold galactic phase, a
key assumption implicitly made above. Possibly this does not happen,
as also suggested in other works (e.g. \citealt{adj+02}), just because
the stellar spheroid is formed too rapidly even compared to an
Eddington limited growth of the MBH from an initial stellar mass up to
$10^8$--$10^9\ms$, which takes $\lesssim 1$\,Gyr. Alternatively, early
MBH growth may be characterized by some duty cycle determined by the
poorly understood physics of gas supply and accretion on to 
protogalaxies.

A proper investigation of the importance of radiative heating on the
MBH--galaxy co-evolution, based on high spatial hydrodynamical
numerical simulations and adopting the specification of the input
spectrum and atomic physics from this work, is now in progress (Ciotti
\& Ostriker, in preparation).


\section*{Acknowledgments}


\appendix
\section{A toy model for the MBH growth}
\label{app_toy}

\subsection{The equations}

The equations describing the evolution of the physical quantities
considered in the present toy-model  are
\beq
\DMgas = \DMinf - \DMstar +\DMrec - \DMbh -\DMesc,
\label{dmgas}
\eeq
\beq
\DMinf = {\Mgal\over\tinf}\exp\left(-{t\over\tinf}\right),
\label{dminf}
\eeq
\beq
\DMstar = {\alstar \Mgas \over\max (\tdyn , \tcool)} - \DMrec,
\label{dmstar}
\eeq
\beq
\DMrec = \int_0^t \DMstar^+(t')\Wstar (t-t')\; dt',
\label{dmrec}
\eeq
where $\DMstar^+$ is the first term on the r.h.s. of equation (\ref{dmstar})
\beq
\DMbh = \DMbhac + \bhstar\DMstar^+, 
\label{dmbh}
\eeq
where
\beq
\DMbhac = \min (\fed\DMedd , \DMbon),
\label{dmbhac}
\eeq
and finally
\beq
\DMesc =\cases{\displaystyle{\Mgas\over\tesc},\quad T \geq\etaesc\Tvir,\cr\cr
                 0,\quad T< \etaesc\Tvir .}
\label{dmesc}
\eeq

\subsection{Input physics}

\subsubsection{Gas equilibrium distribution}

The code is started by assigning the dark-matter halo circular
velocity $\vcirc$ under the assumption of a (singular) isothermal
distribution [see equation~(\ref{rho_gas})], and the quantity $\tdyn$ entering
equation~(\ref{dmstar}) is defined as 
\beq 
\tdyn\equiv {2\pi\Reff\over\vcirc}, 
\eeq 
where $\Reff$ is a characteristic scale-length that could be
identified with the effective radius or the half-mass radius of the
gas and galaxy stellar distribution. For example, the radial trend of
the gas density is obtained by {\it arbitrarily} imposing that all the
gas mass is contained within $2\Reff$, and it is distributed as the
dark matter halo:
\beq 
\rho ={\rhom\over 3} \left ({\Reff\over r}\right)^2.
\label{Arhog}  
\eeq 
The assumption above should not taken too literally, in the sense that
it is only a simple way to obtain a representative value of the mean
gas density $\rhom$: different choices would lead to different values
of $\rhom$ for an assigned total gas mass.  In any case, as a
consequence of our assumption, the {\it mean gas density} within
$\Reff$ is given by
\begin{equation}
\rhom ={3\Mgas\over 8\pi\Reff^3}.
\end{equation}
The (mass-weighted) equilibrium gas temperature $\Tvir$ could be
obtained from the hydrostatic equilibrium and Jeans equations. However
it is well known that a gas distribution as that in equation~(\ref{Arhog})
but untruncated (and so characterized by an infinite mass) would have
the (constant) temperature
\begin{equation}
\Tvir ={\mu\mp\sigma^2\over k}={\mu\mp\vcirc^2\over 2 k},
\label{Atvir}
\end{equation}
where we recall that for the singular isothermal sphere $\sigma^2
=\vcirc^2/2$, where $\sigma$ is the (1-dimensional) halo velocity
dispersion, and $\mu=0.61$.  For simplicity we use equation~(\ref{Atvir})
as an estimate of the equilibrium gas temperature associated with the
density distribution in equation~(\ref{Arhog}).

\subsubsection{Gas cooling time}

The {\it gas mean cooling time} within $\Reff$ is defined as
$\tcool\equiv E/\DEgasC$, where
\begin{equation}
E={3k\rhom T\over 2\mu\mp}
\label{Aien}
\end{equation}
is the gas internal energy per unit volume, 
\begin{equation}
\DEgasC=n_en_p\LT={n_t^2\over 4}\LT =
\left({\rhom\over 2\mu\mp}\right)^2\LT ,
\label{Acool}
\end{equation}
where 
\begin{equation}
\LT={2.18\times 10^{-18}\over T^{0.6826}}+2.706\times 10^{-47}T^{2.976}
\;\ergscm
\end{equation}
(see \citealt{mb78}, CO01).  We thus obtain
\begin{equation}
\tcool ={6\mu\mp k\over\rhom}{T\over\LT}.
\end{equation}

\subsubsection{Stellar physics}

The gas recycled by the evolving stellar population is given by
equation~(\ref{dmrec}).  The modulating (normalized) kernel is given by
\begin{equation}
\Wstar (t)=\Rstar\times {\destar -1\over\tz}
             \left({\tz\over t + \tz}\right)^{\destar}, 
\end{equation}
where we adopt $\Rstar =0.3$, $\destar =1.36$, $\tz = 10^8 \yrs$;
thus, a fraction $\Rstar$ of the stellar mass produced at any time
step is recycled at the end to the galaxy interstellar medium. The
function above is a fairly acceptable fit of the time dependent mass
return rate of a simple, passively evolving stellar population,
adopted in \citet{cdp+91} (hereafter CDPR), CO97, 01.

\subsubsection{MBH accretion physics}

Concerning equations~(\ref{dmbh}) and (\ref{dmbhac}), we define
\begin{equation}
\DMedd\equiv{\Ledd\over\epsilon c^2},
\label{Aedd}
\end{equation}
where $0.001\leq \epsilon\leq 0.1$ and $\Ledd$ is given in
equation~(\ref{ledd}).
The Bondi accretion rate is given by
\begin{equation}
\DMbon = 4\pi\rbon^2\rhobon\csound,
\end{equation}
where $\rbon$ is given by equation~(\ref{rb}) and 
\begin{equation}
\csound^2 =\left({\partial p\over\partial\rho}\right)_{isot}=
{k T\over\mu\mp}.
\end{equation}
An estimate of the gas density at $\rbon$, $\rhobon$, is obtained from
equation~(\ref{Arhog}) evaluated at $r=\rbon$\footnote{Note that the Bondi
accretion rate given by equations above becomes irrelevant when the
gas is colder than the virial temperature, because the Bondi radius is
then larger than the MBH radius of influence. More appropriate in this
case would be to calculate the accretion rate for the conditions at
the influence radius. However, for the adopted gas density profile the
accretion rate will in fact remain unchanged.}.

\subsection{Feedback and gas escape}

In the following two sections we describe two different kinds of {\it
heating} that act on the galactic gas, plus the gas cooling: note that
we are not counting cooling twice, because the cooling function in
equation~(\ref{Acool}) is used to determine the gas cooling time only.
Thus, we determine the gas temperature at each time-step by integrating the
equation for the internal energy per unit volume 
\begin{eqnarray}
\DEgas &=&\DEgasSNw + \DEgasRW + \DEgasAGN - \DEgasC +\DEgasAD +\nonumber\\
       &&3{\DMinf\lambda\vesc^2 -\DMesc\csound^2\over 16\pi\Reff^3},
\label{Adegas}
\end{eqnarray}
where $\lambda$ is a dimensionless parameter ranging from 0.25
(corresponding to virial velocity) to 1; $\DEgasAD$ describes the
adiabatic cooling in case of gas escaping (Sect. A3.4), while
$\DEgasRW$ describes the energy input due to the thermalization of
red-giants wind (Sect. A3.2). As a reference value we use the quantities
within the half mass radius (here identified with $\Reff$) and,
accordingly, the gas internal energy is given by equation~(\ref{Aien}),
while only half of the gas accretion/outflow kinetic energy is
considered.  In particular, in the integration of equation above we
adopt a global fitting formula for the quantity $\DEgasAGN - \DEgasC$
(Sect. A3.3). From equations~(\ref{Aien}) and (\ref{Adegas})
\begin{equation}
T(t+\Delta t)={2\mu\mp E(t+\Delta t)\over 3 k\rhom (t+\Delta t)}.
\end{equation}
At each time-step a check on the attained gas temperature is
performed, and we force the gas temperature to remain above the
minimum value $T_{min}=10^4$ K.  Energy (temperature) integration is
actually the last time-step in the integration cycle. When a new
interation starts, we first check if the new determined temperature is
higher than the escape temperature $\etaesc\Tvir$.  If the condition is not
satisfied (i.e., $T<\etaesc\Tvir$), gas escape is suppressed. On the
contrary, if $T\geq \etaesc\Tvir$, we adopt
\begin{equation}
\tesc\equiv {2\Reff\over\csound}
\end{equation}
in equation (\ref{dmesc}).

Note that an elementary application of the Virial Theorem, coupled
with the definition of escape velocity and under the assumption that
the gas is distributed as the dark matter halo, would predict that its
mass weighted escape temperature correspond to $\etaesc=4$. However,
this is certainly an upper limit, and we expect that feedback effects
will be important at much lower temperatures, say $\simeq 2\Tvir$ or
even less.

\subsubsection{SNII, SNIa, OB and red giants wind heating}

We describe here the first two contributions to the gas heating in
equation~(\ref{Adegas}). The supernova heating $\DEgasSNw=\DEgasII
+\DEgasIa$ is given by the sum of SNII and SNIa energy injection. Due
to the very short life-time of massive stars, the SNII heating is
assumed istantaneous.

We now compute the expected number of SNII events from a new stellar
mass added to the galaxy according to the first term in the r.h.s of
equation~(\ref{dmstar}).  Let $\DMstar^+$ be the total stellar mass
assembled in the unit time, without considering the successive mass
losses.  We adopt a Salpeter IMF distribution, with the standard low
mass cut-off $\Mstari = 0.1\Msun$, while for simplicity we assume an
infinite value for the upper mass:
\begin{equation}
\Psi  = A M^{-(1+x)}=\DMstar^+ (x-1)\Mstari^{x-1} M^{-(1+x)},
\end{equation}
with $x=1.35$.  The number of SNII events per unit time is just given
by the number of stars with $M >\MII= 8\Msun$:
\begin{equation}
\dot\NII=\int_{\MII}^{\infty}\Psi\,dM=
                   \left(1-{1\over x}\right){\DMstar^+\over\Mstari}
                   \left({\Mstari\over\MII}\right)^x.
\end{equation}
The total energy released by the SNII per unit time is then given by
$\dot\NII\Esn$ (where $\Esn =10^{51}\ergs$), while the actual gas
heating is $\etasn\dot\NII\Esn$, where $\etasn$ is an efficiency
factor. {\it Note that the value of the efficiency $\etasn$ is highly
uncertain, depending on the particular status of the ISM. In addition,
in our simple formalism in the factor $\etasn>$ we also consider the
energy injected by stellar winds by the massive OB stars}. In the computed
models we adopt $\etasn=0.5$.

The treatment of SNIa heating is more complicate, being caracterized
by a delayed explosion with respect to the star formation episode.  We
adopt here the same description as in CDPR, CO97, CO01.
Thus, if $\dot\NIa dt'$ is the total number of SNIa that will explode
at $t\geq t'$ as a consequence of the star formation episode at $t'$,
and $\WIa (t-t')$ the normalized explosion rate, one obtains 
\beq
\DEgasIa =\etasn\Esn\int_0^t\dot\NIa (t')\WIa(t-t')dt', 
\eeq 
where for
simplicity we adopted the same efficiency as for SNII. Following
CDPR
\beq 
\WIa(t)={\deIa-1\over\tIa} \left({\tIa\over t + \tIa}\right)^{\deIa}, 
\eeq 
where $\deIa\simeq 1.4-1.6$ and we arbitrarily assume $\tIa\simeq 0.5$ Gyrs.

We do not enter here in the complicate problem of the determination of
the present rate and the total number of SNIa explosions in a given
stellar population: in the present treatment we just assume that for
each of the star formation episodes $\NIa=\thetaIa\NII$. The actual
value of $\thetaIa$ is fixed by (arbitrarily) requiring that the {\it
standard solar proportion} is maintained, i.e., that SNIa provide
$3/4$ of the iron in the ISM, while the remaining $1/4$ is provided by
SNII \citep{rcd+93}. Because each SNIa produces $\simeq 10$
times more Iron that a SNII, then $\thetaIa\simeq 0.3$.

Thus, the gas heating per unit volume within $\Reff$ due to SNII,
stellar winds from OB stars, and SNIa, is given by 
\begin{equation}
\DEgasSNw ={3\etasn\Esn (\dot\NII +\dot\NIa)\over 8\pi\Reff^3},
\label{desn}
\end{equation}
Finally, thermalization of stellar winds emitted by red-giants to the
stellar velocity dispersion ``temperature'' is an important
contribution to the global energy budget of the ISM in early type
galaxies (see, e.g., CDPR). Here we describe this term
as
\begin{equation}
\DEgasRW = {9\DMrec\vcirc^2\over 32\pi\Reff^3}
\end{equation}
i.e., we evaluate the energy input per unit volume inside $\Reff$.

Thus, the characteristic heating time due to the global energy input of 
the evolving stellar population is 
\begin{equation}
\theatSNw ={3k \rhom T/2 \mu\mp\over \DEgasSNw +\DEgasRW}.
\end{equation}

\subsubsection{Adiabatic cooling}

If the gas escape condition is satisfied, the escaping gas not only
carries out kinetic energy, but also decreases its internal energy in
order to expand. We evaluate the cooling due to gas expansion by using
the First Law of Thermodinamics. In particular, the mechanical work of
expansion is given by $-pdV$. During the expansion the gas mass is
conserved, so that $V=\Mgas/\rho$ and $dV=-\Mgas d\rho/\rho^2$. The
variation of internal energy (per unit volume, as required in energy
equation above) due to gas expansion is then given by $p\Mgas
d\rho/(V\rho^2)=pd\ln\rho$. Then, from $\rho\propto r(t)^{-3}$, and by
assuming that the expansion velocity is given by the sound velocity,
we obtain
\begin{equation}
\DEgasAD= -E {\csound\over \Reff}= -2{E\over\tesc},
\end{equation}
where we used the identity $E=3p/2$ and as order of magnitude $r\simeq
2\Reff$. The equation above is integrated with time-splitting. No adiabatic
heating is considered in case of infall, being considered as accretion
of clouds. 

\subsubsection{AGN radiative feedback}
\label{app_agn}

Using {\small XSTAR} \citep{k02} we compute the net volume heating rate
$\dot{E}$ of a cosmic plasma exposed to radiation characterized by the
average quasar SED adopted from SOS. Photoionization equilibrium is
assumed. $\dot{E}$ depends on gas temperature $T$ and on the
ionization parameter $\xi$, which can be determined from
equations~(\ref{xi}) and (\ref{Arhog}) as follows:
\begin{equation}
\xi\equiv {L\over n(r) r^2}={3L\over\rhom\Reff^2},
\label{Acsi}
\end{equation}
where $L=\epsilon c^2\DMbh$ is calculated at each time-step from
equation~(\ref{dmbhac}). Note that from equation~(\ref{Arhog}) $\xi$ is
independent of radius (as far as the gas is optically thin). 

The following furmula provides a good approximation to $\dot{E}$ (all
quantities are expressed in cgs system):
\begin{equation}
\dot E = n^2 (S_1 + S_2 + S_3),
\end{equation}
where $n$ is the H nucleus density (in number),
\begin{equation}
S_1 = -3.8\times 10^{-27}\sqrt{T}
\end{equation}
are the bremsstrahlung losses, 
\begin{equation}
S_2 = 4.1\times 10^{-35} (1.9\times 10^7 -T)\,\xi
\end{equation}
is Compton heating (cooling), and 
\begin{equation}
S_3 = 10^{-23}{a + b\, (\xi/\xi_0)^c\over 1 + (\xi/\xi_0)^c}
\end{equation}
is the sum of photoionization heating, line and recombination
continuum cooling. 
\begin{equation}
a=-{18\over  e^{25 (\log T -4.35)^2}} 
    -{80\over  e^{5.5(\log T -5.2)^2}}
    -{17\over  e^{3.6(\log T -6.5)^2}},
\end{equation}
\begin{equation}
b={1.7\times 10^4\over T^{0.7}}, 
\end{equation}
\begin{equation}
c=1.1-{1.1\over  e^{T/1.8\,10^5}}+{4\times 10^{15}\over T^4}, 
\end{equation} 
and finally
\begin{eqnarray}
\xi_0 &=& {1\over 1.5/\sqrt{T}+1.5\times 10^{12}/\sqrt{T^5}}+\nonumber\\
      &&  {4\times 10^{10}\over T^2}
          \left[1 + {80\over e^{(T-10^4)/1.5\,10^3}}\right].
\end{eqnarray}
The formula above is applicable in the temperature range $10^4\leq
T\leq 3\times 10^7$ independently of the $\xi$ value, except for $T <
2\times 10^4$ when it breaks down at $\xi < 0.01$ (hydrogen becomes neutral),
but then $S_1,S_2,S_3 \to 0$. 

For $T >3\times 10^7$\,K one can use the approximation $\dot E = n^2
(S_1 + S_2)$.

\end{document}